\documentclass[prb,aps,twocolumn,showpacs,amsmath,amssymb]{revtex4-1}

\usepackage[pdftex]{graphicx}
\usepackage{color}
\definecolor{dred}{rgb}{0.7,0.0,0.0}
\usepackage{dcolumn}
\usepackage{bm}
\allowdisplaybreaks 
\usepackage{braket}
\usepackage{ulem} 
\usepackage{hyperref}
\hypersetup{colorlinks=true, linkcolor=blue, citecolor=blue, filecolor=blue, urlcolor=blue, pdftitle=, pdfauthor=, pdfsubject=, pdfkeywords=}

\newcommand{\bs}[1]{\ensuremath{\boldsymbol{#1}}}

\renewcommand{\i}{\text{i}}
\renewcommand{\r}{\text{r}}
\newcommand{\x}{\text{x}}
\newcommand{\y}{\text{y}}
\newcommand{\z}{\text{z}}

\renewcommand{\a}{\text{a}}

\def\ie{{i.e.},\ }

\def\etal{{et al.}}

\newcommand {\bu}{{\bar u}}
\newcommand {\bv}{{\bar v}}

\newcommand{\up}{\uparrow}
\newcommand{\dw}{\downarrow}

\definecolor{orange}{rgb}{1,0.5,0}
\definecolor{black}{rgb}{0,0,0}

\begin{document}

\title{Landau level quantization of Dirac electrons on the sphere}

\author{Martin Greiter} 
\email{greiter@physik.uni-wuerzburg.de}
\affiliation{Institute for Theoretical Physics, 
  University of W\"{u}rzburg, Am Hubland, D-97074 W\"{u}rzburg,
  Germany}

\author{Ronny Thomale} \affiliation{Institute for Theoretical Physics, 
  University of W\"{u}rzburg, Am Hubland, D-97074 W\"{u}rzburg, Germany}

\date{\today}

\begin{abstract}
Interactions in Landau levels can stabilize new phases of matter, such as fractionally quantized Hall states.  Numerical studies of these systems mostly require compact manifolds like the sphere or a torus.  For massive dispersions, a formalism for the lowest Landau level on the sphere was introduced by Haldane [F.D.M.~Haldane, PRL 51, 605 (1983)].  Graphene and surfaces of 3D topological insulators, however, display massless (Dirac) dispersions, and hence require a different description.  We generalize a formalism previously developed for Dirac electrons on the sphere in zero field to include the effect of an external, uniform magnetic field.
\end{abstract}

\maketitle

Progress in theoretical physics has always been achieved through the
interplay of obtaining experimental data with comparing it to the
predictions of the ideas, concepts, and theories suggested to explain
the data.  In earlier periods, the implications of theoretical models
could be explored only through analytic calculations.  During the past
four decades, however, the availability of ever more powerful
computers has significantly reshaped this process.  Among early
highlights were the development of the renormalization group by
Wilson~\cite{wilson75rmp773}, the discovery of universality in the
onset of chaos by Feigenbaum~\cite{feigenbaum78jsp25}, and the
formulation of Laughlin's wave function for fractional quantized Hall
liquids~\cite{laughlin83prl1395}.  Laughlin's discovery is
particularly striking in this context as it was guided by a numerical
experiment~\cite{Laughlin90book}.  Laughlin numerically diagonalized a
system of a few electrons in the lowest Landau level in the open
plane, and observed that the canonical angular momentum of the ground
state jumped by a factor of three upon turning on a strong repulsive
interaction.  The experimental discovery of the effect had inspired
the numerical experiment, and the numerical experiment provided the
crucial hint to the formulation of the theory.  The theory was only
accepted by the community at large after Haldane formulated it on a
sphere~\cite{haldane83prl605}, a geometry without a boundary and hence
without gapless edge modes, and showed that Laughlin's trial state can
be adiabatically connected to the ground state for Coulomb
interactions without closure of the energy gap~\cite{haldane90book}.
A more recent example for the importance of numerical experiments is
the discovery of the topological insulator (TI) as a consequence of
band inversion by Kane and
Mele~\cite{kane-05prl146802,kane-05prl226801}, a phase which was
subsequently realized in HgTe quantum wells~\cite{bernevig-06s1757,koenig-07s766}.

The efficient implementation of numerical experiments often requires
geometries which cannot be realized in a laboratory, such as periodic
boundary conditions (PBCs).  When the underlying lattice plays no role
in the effective model one wishes to study, the simplest geometry
without a boundary is the sphere.  It continues to be of seminal
importance in numerical studies of quantized Hall states and other
states of matter in two dimensional electron gases subject to a
magnetic field.  While most of the work on TIs focusses on the single
particle description of topologically non-trivial band structures, the
most promising avenues to observe topologically non-trivial many body
condensates in this context may be at the surface of a 3D
TI~\cite{vishwanath-13prx011016,metlitski15prb125111}.  The single
particle states on these surfaces are described by a single Dirac
cone, which would be impossible to realize on a lattice due to the
fermion doubling theorem~\cite{niemi-83prl2077}.  Even as a continuum
theory, coupling the electrons minimally to the electromagnetic gauge
field requires an even number of Dirac cones, or an axion term on one
side of the surface~\cite{redlich84prl18}.  In other words, a single
Dirac cone at a surface requires a termination of a topological
insulator~\cite{qi-08prb195424}.  The situation is less intricate in
graphene, where a 2D lattice not embedded in a 3D topological
structure features one Dirac cones per spin and valley degree of
freedom, and hence a total of four cones~\cite{castroneto-09rmp109}.

Regarding the numerical study of interaction effects on surfaces of 3D
TIs, the only work published so far has employed a spherical
geometry~\cite{neupert-15prl017001,durst:arXiv:1603.07676}.  (For
PBCs, the numerics is far more challenging, and the studies performed
so far are unpublished as of yet~\cite{haldanePC}.)  To formulate the
single particle Hilbert space for the single Dirac cone on the sphere,
we employed a formalism introduced earlier by one of
us~\cite{greiter11prb115129} to describe Landau levels (LLs) for
massive electrons on the sphere, which in turn generalized the spinor
coordinate formalism introduced earlier by
Haldane~\cite{haldane83prl605} for the lowest LL.  The magnetic
monopole in the center of the sphere, of monopole charge $2s_0=+1$ for
$\up$ spins and $2s_0=-1$ for $\dw$ spins, emerges from the Berry's
phase associated with rotations of the reference system for the spin.
(In our notation, spin $\up$ and $\dw$ refer to spin directions
normal to the surface of the sphere.)  We obtained the single particle
Hamiltonian,
\begin{align}
  \label{eq:H0withS+S-}
  H=\frac{\hbar v}{R}
  \left(\!\!
    \begin{array}{c@{\hspace{2pt}}c} 0&-S^+\\[2pt]-S^-&0 \end{array}
    \!\!\right)\!,
\end{align}
where the angular momentum operators $S^-$ and $S^+$ effectively act
as LL ``raising'' and ``lowering'' operators on the sphere, $v$ is the
Dirac velocity, and $R$ the radius of the sphere.  This form resembles
the single particle Hamiltonian for Dirac electrons subject to a
uniform magnetic field $\bs{B}=-B\bs{e}_\z$ in the plane,
\begin{align}
  \label{eq:h0witha+a}
  H=\frac{\hbar v\sqrt{2}}{l}
  \left(\!\!
    \begin{array}{c@{\hspace{6pt}}c} 0&\i a^\dagger\,\\[2pt]
     -\i a&0^{\phantom{\dagger}} \end{array}
    \!\!\right)\!,
\end{align}
where $a^\dagger$ and $a$ are Landau level raising and lowering
operators (see Refs.~\onlinecite{Arovas86} or~\onlinecite{Greiter11}
for reviews of the formalism), and $l=\sqrt{\frac{\hbar c}{eB}}$ is
the magnetic length.

In this paper, we first provide a more detailed derivation of
\eqref{eq:H0withS+S-} than space allowed in
Ref.~\onlinecite{neupert-15prl017001}, and second, show that
\eqref{eq:H0withS+S-} also holds in the presence of an external,
radial magnetic field $\bs{B}=B\bs{e}_\r$ supplementing the Berry
flux.  We assume a field strength $B={2b_0 \Phi_0}/{4\pi R^2}$, where
$\Phi_0={2\pi\hbar c}/{e}$ with $e>0$ is the Dirac flux quantum, such
that the total number of Dirac flux quanta through the surface is
$2b_0$.  \emph{The only change due to the field is that the $\up$ and $\dw$
spin components of the spinor $\psi^\lambda_{nm}$},
\begin{align}
  \label{eq:psilambda}
  H\psi^\lambda_{nm}=
  E_n\psi^\lambda_{nm},\ \
  \psi^\lambda_{nm}=\left(\!
    \begin{array}{c}\phi^\up_{nm}\\[2pt] \lambda\phi^\dw_{nm}\end{array}
    \!\right)\!,\
\end{align}
\emph{are given by (massive) LL wave functions~\cite{greiter11prb115129}
corresponding to total magnetic flux $\Phi=(2b_0\pm 1)\Phi_0$ rather
than just $\pm\Phi_0$ through the surface of the sphere}, and the
energies for states in (Dirac) LL $n$ are given by
\begin{align}
  \label{eq:En}
  E_n=\lambda\frac{\hbar v}{R}\sqrt{(2b_0+n)n}
\end{align}
rather than $E_n=\lambda\frac{\hbar v}{R} n$. $\lambda=\pm 1$
distinguishes positive from negative energy solutions.  (Note that
since the level $n=0$ does not exist for the zero field case, $n$ is
shifted by one as compared to the discussion in
Ref.~\onlinecite{neupert-15prl017001}.)

Let us now turn to the details of the derivation.  We consider the Dirac
Hamiltonian
\begin{align}
  \label{eq:Dirac}
  H
  =\hbar v\hat{\bs{n}}
  \left[\left(-\i\nabla-\frac{e}{c}\bs{A}\right)
    \times\bs{\sigma}\right],
\end{align}
where $\hat{\bs{n}}$ is the surface normal, and $\bs{A}$ the vector
potential generating the external magnetic field. Note that the scalar
product with the surface normal ensures a rotationally symmetric form
of the 2D Dirac (surface) Hamiltonian. For the surface
states of a 3D TI, $\bs{\sigma}=(\sigma_\x,\sigma_\y,\sigma_\z)$ is
twice the physical electron spin vector.  For graphene ($\hat{\bs{n}}=\hat{\bs{z}}$), the Pauli
matrices act on the two-dimensional space spanned by the two sites
contained in the unit cell of the hexagonal lattice, usually denoted
as sublattice A and B.  In the case of the TI, the external magnetic
field will also couple to the electron spin via a Zeeman term, but
since this will not give rise to any conceptual difficulties, we will
only address it briefly after the derivation.  In the following, we
set $\hbar=c=1$.

In the absence of the external magnetic field, Imura
\etal~\cite{imura-12prb235119} used the example of a 3D TI to show
that on a sphere with radius $R$, \eqref{eq:Dirac} becomes 
\begin{align}
  \label{eq:H0}
  H_0=\frac{v}{R}\left(\sigma_\x\Lambda_\theta+\sigma_\y\Lambda_\phi\right),
\end{align}
where 
\begin{align}
  \label{eq:lambda}
  \bs{\Lambda}\;=\;-i\left[
    \bs{e}_\varphi\partial_\theta
    -\bs{e}_\theta\frac{1}{\sin\theta}
    \left(\partial_\varphi-\frac{\i}{2}\sigma_\z\cos\theta\right)
  \right]
\end{align}
is the dynamical angular momentum of an electron in the presence of a
magnetic monopole with strength $2\pi\sigma_\z$, and
$(r,\theta,\varphi)$ are spherical coordinates.  The monopole strength
or Berry flux through the sphere is hence $\pm 2\pi$ for $\up$ spins
and $\dw$ spins respectively (\ie spins pointing in the
$\pm\bs{e}_\text{r}$ direction).  The origin of this Berry phase is
easily understood.  Since the coordinate system for our spins (to
which our Pauli matrices $\sigma_\x,\sigma_\y,\sigma_\z$ refer to) is
spanned by $\bs{e}_\theta,\bs{e}_\varphi,\bs{e}_\text{r}$, it will
rotate as the electron is taken around the sphere.  For general
trajectories, the Berry phase generated by this rotation is given by
$\pm\frac{1}{2}$ times the solid angle subtended by the trajectory.
Formally, this phase is generated by a monopole with strength
$\pm2\pi$ at the origin for $\up$ and $\dw$ spins, respectively.
Substitution of \eqref{eq:lambda} into \eqref{eq:H0} yields
\begin{align}
  \label{eq:h}
  H&=\frac{v}{R}\,h,\quad
  h=\!\left(\!\!
    \begin{array}{cc}
      0&h^+\\h^-&0
    \end{array}\!\!\right)\!,
\end{align}
with 
\begin{align}
  \label{eq:h0+-}
  h^\pm = h_0^\pm = {\mp\left(\partial_\theta+\frac{1}{2}\cot\theta\right)
    +\frac{\i\partial_\varphi}{\sin\theta}}.
\end{align} 

Even though Imura \etal~\cite{imura-12prb235119} derived \eqref{eq:h}
with \eqref{eq:h0+-} discussing the surface termination of a 3D TI, it
is by no means specific to this setting, as \emph{the Berry phase is a
general property of the Dirac Hamiltonian on a curved surface.}  To
illustrate this point, we will derive \eqref{eq:h} with
\eqref{eq:h0+-} now directly from \eqref{eq:Dirac} with $\bs{A}=0$.

On a sphere with fixed radius $R$, the nabla operator in spherical
coordinates reads
\begin{align}
  \label{eq:nabla}
  \nabla=\frac{1}{R}\left(\bs{e}_\theta\partial_\theta
    +\bs{e}_\varphi\frac{\partial_\varphi}{\sin\theta}\right).
\end{align}
This form, however, is not suited for direct substitution into
\eqref{eq:Dirac}, since $-\i\nabla$ has to be hermitian, while 
\begin{align}
  \label{eq:hermconj}
  (\partial_\theta)^\dagger=-(\partial_\theta+\cot\theta),\quad
  (\partial_\varphi)^\dagger=-\partial_\varphi.
\end{align}
(The solid angle measure $d\Omega=d\theta d\phi\sin\theta$ gives rise
to the $\cot\theta$ term when we go from
$\psi_\a^*\partial_\theta\psi_b$ to $-(\partial_\theta
\psi_\a^*)\psi_b$ via partial integration.)
If we then substitute the hermitian combination
  $\frac{1}{2}\!\left((-\i\nabla)+(-\i\nabla)^\dagger\right)$
and
$\bs{\sigma}=
\bs{e}_\theta\sigma_\x+\bs{e}_\varphi\sigma_\y+\bs{e}_\text{r}\sigma_\z$
into \eqref{eq:Dirac}, we obtain \eqref{eq:h} with \eqref{eq:h0+-}.
 
To include the external magnetic field, we choose the latitudinal
gauge
\begin{equation}
  \label{eq:gauge}
  \bs{A}=-\bs{e}_\varphi\frac{b_0}{eR}\cot\theta.
\end{equation}
The singularities of $\bs{B}=\nabla\times\bs{A}$ at the poles are
without physical significance.  They describe infinitly thin solenoids
admitting flux $b_0\Phi_0$ each, exist for the Berry connection as
well, and reflect our inability to formulate true magnetic monopoles.

Substitution of \eqref{eq:gauge} into \eqref{eq:Dirac} yields \eqref{eq:h}
with 
\begin{align}
  \label{eq:h+-}
  h^\pm &= h_0^\pm + b_0\cot\theta\nonumber\\[10pt]
  &= {\mp\,\partial_\theta + \left(b_0\mp\frac{1}{2}\right)\cot\theta
    +\frac{\i\partial_\varphi}{\sin\theta}}.
\end{align} 
As in the zero field case, \eqref{eq:h} with \eqref{eq:h+-} describes a ``Dirac Hamiltonian''
in the sense that
\begin{align}
  \label{eq:h^2}
  h^2&=\left(\!
    \begin{array}{cc}
      h^+h^-&0 \\0&h^-h^+
    \end{array}\!\!\right)\nonumber\\[5pt]
  &=\left(\!
    \begin{array}{cc}
      \bs{\Lambda}^2_{s_0}+s_0\left|_{s_0=b_0+\frac{1}{2}}\right.&0 \\[10pt]
      0&\bs{\Lambda}^2_{s_0}-s_0\left|_{s_0=b_0-\frac{1}{2}}\right.
    \end{array}\!\right)
\end{align}
is diagonal.  Apart from an overall numerical factor,
\begin{align}
  \label{eq:qhslambda^2}
  \bs{\Lambda}^2_{s_0}=
     -\frac{1}{\sin\theta}\partial_\theta
     \left(\sin\theta\,\partial_\theta\right)
     -\frac{1}{\sin^2\theta}
     \left(\partial_\varphi-\i s_0\cos\theta\right)^2
\end{align}
is the Hamiltonian of a massive electron moving on a sphere with a
monopole of strength $4\pi s_0$ in the center~\cite{haldane83prl605}.
The LLs for massive electrons on the sphere are spanned by two
mutually commuting SU(2) algebras~\cite{greiter11prb115129}, one for
the cyclotron momentum ($\bs{S}$) and one for the guiding center
momentum ($\bs{L}$). The Casimir of both is given by
$\bs{L}^2=\bs{S}^2=s(s+1)$, where $s=|s_0|+n$ and $n=0,1,\ldots$ is
the LL index for massive electrons.

With
$\bs{\Lambda}^2=\bs{L}^2-s_0^2$, we obtain
\begin{align}
  \label{eq:lambda^2pm}
  \bs{\Lambda}^2_{s_0}\pm s_0\left|_{s_0=b_0\pm\frac{1}{2}}\right.=
  \left\{\begin{array}{ll}
    (2b_0+n_{\up}+1)(n_{\up}+1),\\[5pt] 
    (2b_0+n_{\dw})n_{\dw},
    \end{array}\right.
\end{align}
for the diagonal elements of $\up$ and $\dw$ spins in \eqref{eq:h^2}.
The $\up$ spin components $\phi^\up_{nm}$ are hence described by
massive LL wave functions in level $n_{\up}=n-1$ if the $\dw$ spin
components $\phi^\dw_{nm}$ are described by massive LL wave functions
in level $n_{\dw}=n$, with $s=b_0+n-\frac{1}{2}$ for both.  The
eigenvalues of $h^2$ are given by $\varepsilon_n^2=(2b_0+n)n$.

In terms of the spinor coordinates 
\begin{align}
  \label{eq:uv} \textstyle
  u=\cos\frac{\theta}{2} e^{i\frac{\varphi}{2}},\
  v=\sin\frac{\theta}{2} e^{-i\frac{\varphi}{2}},
\end{align}
introduced by Haldane \cite{haldane83prl605}, and their complex
conjugates $\bar u$, $\bar v$, 
%
%
\begin{align}
  \label{eq:Sdef}
  S^x + iS^y = S^+&=u\partial_\bv-v\partial_\bu,\nonumber\\ 
  S^x - iS^y = S^-&=\bv\partial_u-\bu\partial_v,\\\nonumber
  S^\z &= \textstyle\frac{1}{2}
  (u\partial_u + v\partial_v - \bu\partial_\bu - \bv\partial_\bv),\\[2pt]
  \label{eq:Ldef}
  L^x + iL^y = L^+&=u\partial_v-\bv\partial_\bu,\nonumber\\ 
  L^x - iL^y = L^-&=v\partial_u-\bu\partial_\bv,\\\nonumber
  L^\z &= \textstyle\frac{1}{2}
  (u\partial_u - v\partial_v - \bu\partial_\bu + \bv\partial_\bv).
\end{align}
The physical Hilbert space is restricted to states with $S^\z$
eigenvalue $s_0$
~\cite{greiter11prb115129}. 
For our spin component wave functions, this restriction reads
\begin{align}
  \label{eq:SzEv}
  \textstyle
  S^\z\phi^\up_{nm}=\left(b_0+\frac{1}{2}\right)\phi^\up_{nm},\ \ 
  S^\z\phi^\dw_{nm}=\left(b_0-\frac{1}{2}\right)\phi^\dw_{nm}.
\end{align}
The greatly simplifying observation is now that for massive LL wave
functions subject to \eqref{eq:SzEv},
\begin{align}
  \label{eq:h+=-S+}
  h^+\phi^\dw_{nm}=-S^+\phi^\dw_{nm},\ \
  h^-\phi^\up_{nm}=-S^-\phi^\up_{nm},
\end{align}
and hence that 
\begin{align}
  \label{eq:hwithS+S-}
  h=\left(\!\!
    \begin{array}{c@{\hspace{2pt}}c} 0&-S^+\\[2pt]-S^-&0 \end{array}
    \!\!\right)\!.
\end{align}

We now verify the first equation in \eqref{eq:h+=-S+} by explicit
evaluation of $h^+\phi^\dw_{nm}$.  $\phi^\dw_{nm}$ has to take the
form of a massive Landau level wave function~\cite{greiter11prb115129}
\begin{align}
  \nonumber
  \phi^\dw_{nm}\sim (L^-)^{s-m}(S^-)^n u^{2s}
  \sim(L^-)^{s-m}\,{\bar v}^n u^{2s-n}
\end{align}
with $s=b_0+n-\frac{1}{2}$.  Upon expansion we obtain terms of the form
\begin{align}
  \chi_q^{\dw}&=v^{s-m-q}\, {\bar v}^{n-q}\, {\bar u}^q\, u^{s-n+m+q}
  \nonumber\\[2pt]\nonumber
  &={\textstyle\left(\sin\frac{\theta}{2}\right)^{s+n-m-2q}
  \left(\cos\frac{\theta}{2}\right)^{s-n+m+2q}}\, e^{\i m\varphi}
\end{align}
with $q=0,\ldots ,s-m$.  Rewriting \eqref{eq:h+-} as
\begin{align}
  \nonumber
  h^\pm =\mp\,\partial_\theta 
  \textstyle
  +\left(b_0\mp\frac{1}{2}+\i\partial_\varphi\right)\cot\frac{\theta}{2}
  -\left(b_0\mp\frac{1}{2}-\i\partial_\varphi\right)\tan\frac{\theta}{2},
\end{align} 
we easily find 
\begin{align}
  \nonumber\textstyle
  h^+\chi_q^{\dw}=\left[
    -(n-q)\cot\frac{\theta}{2}+q\tan\frac{\theta}{2}
    \right]\chi_q^{\dw},
\end{align}
which is equal to $-S^+\chi_q^{\dw}$.  The second equation in
\eqref{eq:h+=-S+} is shown along the same lines.

The Dirac property of $h$, the eigenvalues of $h^2$, the massive
LL form of the component wave functions of the Dirac spinor,
and finally \eqref{eq:hwithS+S-} imply 
\begin{align}
  \label{eq:psilambda}
  h\psi^\lambda_{nm}=\lambda\sqrt{(2b_0+n)n}\,\psi^\lambda_{nm},\ \
  \psi^\lambda_{nm}=\left(\!
    \begin{array}{c}\phi^\up_{nm}\\[2pt] \lambda\phi^\dw_{nm}\end{array}
    \!\right)\!,\
\end{align}
where $\lambda=\pm 1$ distinguishes positive and negativ energy
solutions, and $m$ is the eigenvalue of $L^\z$.  The (only relatively
normalized) component wave functions are given by 
\begin{align}
  \label{eq:phiup}
  \phi^\up_{nm}&=
  \sqrt{n}\; (L^-)^{s-m}\,{\bar v}^{n-1} u^{2s+1-n},\\[5pt]\label{eq:phidw}
  \phi^\dw_{nm}&=
  -\sqrt{2b_0+n}\;(L^-)^{s-m}\,{\bar v}^n u^{2s-n},
\end{align}
where $s=b_0+n-\frac{1}{2}$ and $m=-s,-s+1,\ldots ,s$.  The degeneracy
in each Dirac LL is hence $2s+1=2(b_0+n)$. The level $n=0$ with
dimensionless energy $\varepsilon_0=0$ is completely spin polarized,
with the spins aligned in the direction of $-\bs{B}$ as
$\phi^\up_{0m}=0$; this level does not exist for the zero field case
elaborated in Ref.~\onlinecite{neupert-15prl017001}.  In all other levels,
the single particle states have equal amplitudes for $\up$ and $\dw$
spins.

To gain further insight into the single particle wave functions,
consider the fully normalized spinors for $m=s$, \ie for states
localized at the north pole of the sphere,
\begin{align}
  \label{eq:psinorth}
  \psi^\lambda_{ns} =\sqrt{\frac{1}{2}\!
    \left(\!\begin{array}{c}2(b_0+n)\\n \end{array}\!\right)}&\cdot
  {\textstyle\left(\sin\frac{\theta}{2}\right)^{n-1}
    \left(\cos\frac{\theta}{2}\right)^{2b_0+n-1}}
  \nonumber\\[4pt] &\hspace{-30pt}\cdot\,
  \left(\!
    \begin{array}{c}\sqrt{n}\;\cos\frac{\theta}{2}\\[8pt]
      -\lambda\sqrt{2b_0+n}\;\sin\frac{\theta}{2}\end{array}
    \!\right)\!.
\end{align}
We see that for $n\ne 0$, the spins are aligned with the magnetic
field at the pole, and then turn in the $\sigma_\x,\sigma_\z$ plane
spanned by $\bs{e}_\theta$ and $\bs{e}_\text{r}$ until they point in
the direction opposing the magnetic field far away from the pole.
Almost all the amplitude is contained in narrow rings, which have
their maximal amplitudes at
\begin{align}
  \label{eq:thetamax}
  {\textstyle\left(\tan\frac{\theta}{2}\right)^4}
  =\frac{n(n-1)}{(2b_0+n)(2b_0+n-1)}.
\end{align}
This concludes our derivation of LL quantization for Dirac fermions on
the sphere as applicable to graphene.

For the surfaces of 3D TIs, the spin in \eqref{eq:Dirac} is the
physical electron spin, which also couples to the magnetic field via a
Zeeman term,
\begin{align}
  \label{eq:HZeeman}
  H_{\text{B}}=-\frac{1}{2}g_{\text{s}}\mu_{\text{B}}B\sigma_\z,
\end{align}
where $\mu_{\text{B}}$ is the Bohr magneton, and $g_{\text{s}}$ the
Land\'e $g$-factor.  Even though it is only a small correction in
actual TI surface states, we briefly address it here.  For $n=0$, the
Dirac LL is completely spin polarized and $\psi^\lambda_{nm}$ with
\eqref{eq:phiup} and \eqref{eq:phidw} is an eigenstate of
\eqref{eq:HZeeman}.  The energy is given by
\begin{align}
  \label{eq:EZeeman0}
  E_0=-\frac{1}{2}g_{\text{s}}\mu_{\text{B}}B.
\end{align}
For $n\ne 0$, let the spinor $(\alpha,\beta)^{\text{T}}$
refer to a combination of positive and negative energy solutions as
given by \eqref{eq:psilambda} with \eqref{eq:phiup} and
\eqref{eq:phidw}, $\alpha\psi^+_{nm}+\beta\psi^-_{nm}$.  In this
two-dimensional space, the total Hamiltonian including
\eqref{eq:HZeeman} is given by
\begin{align}
  \label{eq:tildeH}
  \tilde H=\frac{\hbar v}{R}\varepsilon_n\sigma_\z
  -\frac{1}{2}g_{\text{s}}\mu_{\text{B}}B\;\!\sigma_\x,
\end{align}
where $\varepsilon_n=\sqrt{(2b_0+n)n}$ is the absolute value of the
dimensionless energy in the absence of the Zeeman field.  This is
again a ``Dirac Hamiltonian'' in the sense that the square is
diagonal, which allows us to read off the energies
\begin{align}
  \label{eq:tildeE_n}
  \tilde{E}_n=\textstyle\lambda\sqrt{
    \left(\frac{\hbar v}{R}\right)^2\!\, (2b_0+n)n +
    \left(\frac{1}{2}g_{\text{s}}\mu_{\text{B}}B \right)^2},
\end{align}
where $\lambda=\pm 1$ again distinguishes positive and negative energy
solutions. The eigenstates of \eqref{eq:tildeH} are given by
\begin{align}
  \label{eq:tildepsi+}
  \tilde{\psi}^+_{nm} &\textstyle =\left(\frac{\hbar
      v}{R}\varepsilon_n+|\tilde{E}_n|\right) \psi^+_{nm} -
  \left(\frac{1}{2}g_{\text{s}}\mu_{\text{B}}B\right) \psi^-_{nm},
  \\[2pt]\label{eq:tildepsi-}
  \tilde{\psi}^-_{nm}
  &\textstyle
  =\left(\frac{1}{2}g_{\text{s}}\mu_{\text{B}}B\right)\psi^+_{nm} +
  \left(\frac{\hbar v}{R}\varepsilon_n+|\tilde{E}_n|\right)
  \psi^-_{nm}.
\end{align}
The Zeeman term hence yields only a small mixing of the positive and
negative energy solution of the Dirac LLs \eqref{eq:psilambda} with \eqref{eq:phiup} and \eqref{eq:phidw}.

In conclusion, we have presented a formalism for Landau level
quantization of Dirac electrons in the spherical geometry.  The
formalism is largely identical to the formalism we introduced for
Dirac electrons without an external magnetic field in
Ref.~\onlinecite{neupert-15prl017001}, where the issue of Laudau level
quantization arose due to the Berry connection associated with the
coupling of the Dirac spinor to the curvature of the sphere.  Since
the formalism is not limited to either zero field nor to surface
states of 3D TIs, but applies to any other 2D system with Dirac cones
such as graphene, the importance of it goes way beyond the
immediate applications studied in Ref.~\onlinecite{neupert-15prl017001}.

\emph{Note added.}---After this work was completed, we became aware of
  previous articles by Jellal~\cite{jellal08npb361} and
  Schliemann~\cite{schliemann08prb195426}, which also address the
  problem of Dirac electrons in a magnetic field on a sphere.

\begin{acknowledgments}
  This work was supported by the DFG through SFB 1170 ToCoTronics
  and the European Research Council through the grant TOPOLECTRICS
  (ERC-StG-Thomale-336012).
\end{acknowledgments}

\end{document}